\documentclass{article}
\usepackage{amssymb}
\usepackage{amsmath}
\usepackage{graphicx}

\begin{document}

\begin{center}
\Large\textbf{DIFFERENT SHAPES OF IMPURITY CONCENTRATION PROFILES
FORMED BY LONG-RANGE INTERSTITIAL MIGRATION}
\\[2ex]
\normalsize
\end{center}

\begin{center}
\textbf{O. I. Velichko}

\bigskip


{\it E-mail address (Oleg Velichko):} velichkomail@gmail.com
\end{center}

\textit{Abstract.} A model of interstitial impurity migration is
proposed which explains the redistribution of ion-implanted boron
in low-temperature annealing of nonamorphized silicon layers. It
is supposed that nonequilibrium boron interstitials are generated
either in the course of ion implantation or at the initial stage
of thermal treatment and that they migrate inward and to the
surface of a semiconductor in the basic stage of annealing. It is
shown that the form of the ``tail'' in the boron profile with the
logarithmic concentration axis changes from a straight line if the
average lifetime of impurity interstitials is substantially
shorter than the annealing duration to that bending upwards for
increasing lifetime.

The calculated impurity concentration profiles are in excellent
agreement with the experimental data describing the redistribution
of implanted boron for low-temperature annealing at 750 Celsius
degrees for 1 h and at 800 Celsius degrees for 35 min.
Simultaneously, the experimental phenomenon of incomplete
electrical activation of boron atoms in the ``tail'' region is
naturally explained.

\section{Introduction}
The influence of the impurity diffusion mechanism on the form of
dopant distribution after annealing of $\delta $-doped layers has
been investigated in \cite{Cowern-90,Cowern-91}. The authors
obtained a one-dimensional analytical solution of the system of
equations that includes the conservation law for the immobile
component (substitutionally dissolved impurity atoms) and the
diffusion equation for the mobile impurity component
\cite{Cowern-90}. It was supposed that exchange of impurity atoms
occurs between mobile and immobile species. It was also supposed
that the initial impurity distribution is described by the $\delta
$-function and that the generation rate of mobile impurity
interstitials is proportional to the concentration of
substitutionally dissolved impurity atoms. It follows from the
solution obtained that the extended ``tail'', formed in the region
of low impurity concentration of dopant profile after annealing,
represents a straight line for the logarithmic concentration axis
if the impurity atom migrates once. On the other hand, in the case
of multiple events of migration of the interstitial impurity atom,
the Gaussian distribution describes the shape of the impurity
profile in the ``tail'' region.

As follows from the experimental data, after annealing the
impurity concentration profiles can be similar to the Gaussian
distribution. On the other hand, many of the measured impurity
profiles, especially after low-temperature treatment, are
characterized by an extended ``tail'' which looks like a straight
line if the concentration axis is logarithmic (see, for example,
the review of the experimental data in \cite{Velichko-12}).
Therefore, it was proposed in \cite{Cowern-90,Cowern-91} that the
microscopic diffusion mechanism of impurity atoms be determined
based on the form of the dopant distribution after annealing. It
is worth noting that in \cite{Cowern-90,Cowern-91} continuous
generation of mobile species in the region of high impurity
concentration was supposed in the case where an impurity atom
migrates once. In \cite{Velichko-07,Velichko-2011} analytical
solutions were obtained for the initial impurity distributions
that, in contrast to \cite{Cowern-90,Cowern-91}, represent an
uniform doped layer or the Gaussian distribution. It was shown
that the ``tails'' also represent a straight line just as for the
initial distribution described by $\delta $-function
\cite{Cowern-90,Cowern-91}.

At present, the method of ion implantation with low budget
annealing such as, for example, high-temperature spice annealing
or low-temperature long-term thermal treatment, is widely used for
fabricating modern semiconductor devices and integrated
microcircuits.  During ion implantation, a fraction of the
implanted ions can occupy an interstitial position and participate
in the interstitial migration at subsequent annealing. It is
expected that this phenomenon will be most pronounced for
implantation of light ions, for example, boron ions, and also for
implantation of the medium doses of heavier ions
($\sim$10${}^{14}$ cm${}^{-2}$) when the maximal concentration of
impurity atoms is below the solubility limit but higher than the
intrinsic carrier concentration at the annealing temperature
\cite{Huang-97}. Indeed, in these two cases one does not observe
the amorphization of an implanted layer. As a result, annealing of
defects generated by ion implantation occurs during the total
thermal treatment, instead of the solid phase recrystallization of
the created amorphous layer at the initial stage of the thermal
processing. Indeed, it is well known that during solid phase
epitaxial regrowth (SPER) there occurs annealing of the main part
of defects created by ion implantation. In addition, almost all
implanted impurity atoms go over into the substitutional position,
even for the concentration above the solubility limit. It will be
expected that in the absence of SPER the impurity atoms that
remained in the interstitial position will migrate to a
significant distance before they become substitutionally
dissolved. It is worth noting that in the case of low-temperature
treatment with duration of ten minutes and longer, a significant
number of the impurity interstitials can also be generated at the
initial stage of thermal processing due to annealing or
rearrangement of different kind of defects incorporating impurity
atoms.

Therefore, in contrast to the case of continuous generation of
impurity interstitials during annealing described in
\cite{Cowern-90,Cowern-91,Velichko-07,Velichko-2011}, the purpose
of this work is to investigate the characteristic features of
ion-implanted impurity redistribution that can appear due to
diffusion of the previously formed interstitial impurity atoms.

\section{The system of the equations describing interstitial diffusion}
Let us assume that the substitutionally dissolved impurity atoms,
as well as the impurity atoms incorporated into clusters,
precipitates, and radiation defects are immobile and do not
participate in diffusion due to the low annealing temperature. As
the formation of extended ``tails'' due to the migration of
impurity interstitials occurs in the region of low impurity
concentration $C\le n_{i} $, we neglect the influence of the
built-in electric field on the diffusion process. Here $C$ is the
concentration of substitutionally dissolved impurity atoms and
$n_{i}$ is the intrinsic carrier concentration. Then, the system
of equations that was proposed in \cite{Velichko-83,Velichko-88}
and that describes migration of nonequilibrium impurity
interstitial atoms and their subsequent transition to the
substitutional position can be presented in the form given below.

\noindent \textbf{1. The conservation law for substitutionally
dissolved impurity atoms:}

\begin{equation} \label{Conservation law}
\frac{\partial C(x,t)}{\partial t} =\frac{C^{AI} (x,t)}{\tau ^{AI} } \,  .
\end{equation}

\noindent \textbf{2. The equation of diffusion for nonequilibrium
impurity interstitials:}

\begin{equation} \label{Nonequilibrium impurity interstitials}
\frac{\partial C^{AI} }{\partial t} =d^{AI} \frac{\partial ^{2}
C^{AI}}{\partial x^{2} } -\frac{C^{AI} }{\tau^{AI} } +G^{AIR}
(x,t) ,
\end{equation}

\noindent where $C^{AI}$ is the concentration of nonequilibrium
interstitial impurity atoms (IIA); $d^{AI} $ and $\tau ^{AI} $ are
the diffusivity and average lifetime of these impurity
interstitials, respectively; $G^{AIR} $ is the generation rate of
interstitial impurity atoms per unit volume due to the annealing
of implantation defects and  rearrangement of clusters
(precipitates). It is worth noting that the total concentration of
impurity atoms $C^{T} $ includes concentrations of the
substitutionally dissolved impurity atoms, impurity atoms
incorporated into clusters, precipitates, and radiation defects,
as well as the concentration of impurity interstitial atoms
$C^{AI}$.

It is well known that interstitial impurity atoms can become
substitutionally dissolved due to the recombination with vacancies
(the so-called the Frank-Turnbull diffusion mechanism
\cite{Frank-56}). In addition, a migrating interstitial impurity
atom can kick-out the host atom and occupy a substitutional
position (the reverse Watkins effect). Then, the average lifetime
of IIA can be presented in the following form:

\begin{equation} \label{tau}
\frac{1}{\tau ^{AI} } =k^{AIV} C^{V} +k^{AI}  ,
\end{equation}

\noindent where $k^{AIV}$ is the effective coefficient of
recombination of interstitial impurity atoms with vacancies;
$C^{V} $ is the vacancy concentration; $k^{AI} $ is the
coefficient describing transition of the interstitial impurity
atom to the substitutional position due to the kick-out mechanism.

It is worth noting that this paper investigates an extreme case of
the redistribution of implanted atoms occupying interstitial
position immediately after implantation. Therefore, the generation
term $G^{AIR}=0$. Then, Eqs. \eqref{Conservation law} and
\eqref{Nonequilibrium impurity interstitials} are independent if
the vacancy distribution is uniform ($C^{V} \left(x,t\right)={\rm
const}$). If $d^{AI}$ is also constant, one can obtain an
analytical solution of Eq. \eqref{Nonequilibrium impurity
interstitials}.

\section{Analytical solution of nonstationary diffusion equation
for interstitial impurity atoms} An analytical solution of Eq.
\eqref{Nonequilibrium impurity interstitials} can be obtained
within the framework of the Green function approach
\cite{Tichonov-90,Budak-88,Duffy-01}. For this purpose let us
present Eq. \eqref{Nonequilibrium impurity interstitials} in the
following form:

\begin{equation} \label{Normalized equation}
\frac{\partial \, C^{AI} }{\partial \, \theta } =l_{AI}^{2}
\frac{\partial ^{2} C^{AI} }{\partial \, x^{2} } -C^{AI}  ,
\end{equation}

\noindent where $l_{AI} =\sqrt{d^{AI} \tau ^{AI}} $ is the average
migration length of impurity interstitials; $\theta ={t
\mathord{\left/{\vphantom{t \tau ^{AI}
}}\right.\kern-\nulldelimiterspace} \tau ^{AI} } $ is the relative
time.

Let us also assume that the distribution of nonequilibrium
impurity interstitials after implantation and after rapid
annealing of implantation defects incorporating impurity atoms is
proportional to the distribution of the implanted atoms which can
be approximately described by the Gaussian distribution. Then, the
initial condition for impurity interstitials can be presented in
the following form:

\begin{equation} \label{Initial condition}
C_{0}^{AI} (x)=C^{AI} (x,\; 0)=C_{m}^{AI} \exp
\left[-\frac{\left(x-R_{p} \right)^{\, 2}}{2\Delta R_{p} ^{2}}
\right] ,
\end{equation}

\noindent where $C_{m}^{AI}$ is the maximal concentration of
interstitial impurity atoms after implantation or at the initial
stage of annealing; $R_{p} $ and $\Delta R_{p} $ are the average
projective range of implanted ions and the straggling of the
projective range, respectively.

Let us obtain an analytical solution of the diffusion equation for
nonequilibrium impurity interstitials \eqref{Normalized equation}
in one-dimensional (1D) semiinfinite domain $\left[0,\; +\infty
\right]$. For the sake of simplicity, the Dirichlet boundary
condition can be imposed on the surface and in the bulk of a
semiconductor:

\begin{equation} \label{Dirichlet}
C^{AI} (0,\; \theta )=C_{S}^{AI}  , \qquad \qquad \qquad  C^{AI}
(+\infty ,\; \theta )=0 ,
\end{equation}

\noindent where $C_{S}^{AI}$ is the concentration of interstitial
impurity atoms on the surface of a semiconductor.

For this purpose, let us consider Eq. \eqref{Normalized equation}
in the infinite domain $\left[-\infty ,\; +\infty \right]$ and use
separation of variables \cite{Tichonov-90,Farlow-93} assuming that
the required solution is a product of two functions, one of which
depends only on the relative time $\theta $, whereas the other is
the function of only the spatial coordinate $x$, i.e., $C^{AI}
(x,\theta )=T(\theta )X(x)$. Then, $C_{\theta }^{AI} =T'X$,
$C_{xx}^{AI} =TX''$ and Eq. \eqref{Normalized equation} transforms
into the equation

\begin{equation} \label{TX}
T'X=l_{AI}^{2} TX''-TX .
\end{equation}

Dividing the right- and left-hand sides of Eq. \eqref{TX} by
$\left(l_{AI}^{2} TX\right)$, we obtain the following equation:

\begin{equation} \label{Separation}
\frac{T'}{l_{AI}^{2}T} =\frac{X''}{X} -\frac{1}{l_{AI}^{2} }
=-\lambda ^{*2} ,
\end{equation}

\noindent where $\lambda ^{*2} $ is the parameter of division that
depends neither on $x$, nor on $\theta$, i. e., represents a
certain constant. Using this property of $\lambda ^{*2}$, one can
obtain the following system of independent equations:

\begin{equation} \label{Time}
\frac{T'}{T} =-l_{AI}^{2} \lambda ^{*2},
\end{equation}

\begin{equation} \label{Coordinate}
X''+\lambda ^{2} X=0 ,
\end{equation}

\noindent where $\lambda ^{2} =\lambda ^{*2} -{1
\mathord{\left/{\vphantom{1 l_{AI}^{2}
}}\right.\kern-\nulldelimiterspace} l_{AI}^{2}} $ .

The solution of Eq. \eqref{Time} has the following form:

\begin{equation} \label{Time Solution}
T(\theta )=T_{0} \exp \left[\, -l_{AI}^{2} \lambda ^{*2} \theta \right] ,
\end{equation}

\noindent where $T_{0}$ is the value of the function $T(\theta)$
for the initial moment of annealing.

The solution of homogeneous equation \eqref{Coordinate} can be
obtained by employing the standard substitution $X(x)=A_{0}
e^{\alpha {\kern 1pt} x} $, where $A_{0} $ and $\alpha $ are some
constants. Based on the obtained solutions of Eqs. \eqref{Time}
and \eqref{Coordinate} and using the procedure described in
\cite{Tichonov-90}, it is possible to find the solution of the
boundary value problem \eqref{Normalized equation}, \eqref{Initial
condition}, and \eqref{Dirichlet} for zero boundary condition on
the surface of a semiconductor, $C_{S}^{AI} =0$. This solution has
the following form:

\begin{equation} \label{Solution}
C_{ZeS}^{AI} (x,\; \theta )=\int _{0}^{\infty }G(x,\xi ,\theta )
\, C_{0}^{AI} (\xi )d\xi  ,
\end{equation}

\noindent where $G(x,\xi ,\theta )$ is the Green function

\begin{equation} \label{Green function}
G(x,\xi ,\theta )=\frac{1}{2\sqrt{\pi l_{AI}^{2} \theta } }
\left\{\exp \left[-\frac{(x-\xi )^{2} }{4l_{AI}^{2} \theta }
\right]-\exp \left[-\frac{(x+\xi )^{2} }{4l_{AI}^{2} \theta }
\right]\right\} .
\end{equation}

To obtain the required distribution of the concentration of
impurity interstitials for nonzero condition on the surface of a
semiconductor, it is necessary to combine expression
\eqref{Solution} with the well-known solution of the boundary
value problem \eqref{Normalized equation}, \eqref{Dirichlet} with
zero initial condition \eqref{Initial condition}:

\begin{equation} \label{Hom}
C^{AI} (x,\; \theta )=C_{ZeS}^{AI} (x,\; \theta )+C_{S}^{AI}
\left[1-{\rm erf}(\frac{x}{2\sqrt{l_{AI}^{2} \theta } } )\right] .
\end{equation}

A similar procedure can be used to obtain a solution for the
reflecting boundary condition on the surface and zero Dirichlet
boundary condition in the bulk of a semiconductor:

\begin{equation} \label{Nonflow}
\left. \frac{\partial \, C^{AI} }{\partial \, x} \right|_{x=0} =0,
\qquad \qquad \qquad  C^{AI} (+\infty ,\; \theta )=0 .
\end{equation}

This solution has the following form:

\begin{equation} \label{Solution Nonflow}
C^{AI} (x,\; \theta )=\int _{0}^{\infty }G(x,\xi ,\theta) \,
C_{0}^{AI} (\xi )d\xi  ,
\end{equation}

\noindent where the Green function is

\begin{equation} \label{Green function Nonflow}
G(x,\xi ,\theta )=\frac{1}{2\sqrt{\pi l_{AI}^{2} \theta}}
\left\{\exp \left[-\frac{(x-\xi )^{2} }{4l_{AI}^{2} \theta}
\right]+\exp \left[-\frac{(x+\xi )^{2} }{4l_{AI}^{2} \theta}
\right]\right\} .
\end{equation}

The explicit expressions for the integrals in \eqref{Solution} and
\eqref{Nonflow} can be obtained using the modern software allowing
symbolic computation, such as ``Mathematica'' \cite{Mathematica},
``Mathcad'' \cite{Mathcad}, and others. Unfortunately, the
expressions obtained are too lengthy to be presented in this
paper.

The total concentration of impurity atoms after annealing can be
calculated by Eq. \eqref{Conservation law}. Integrating this
equation, we obtain

\begin{equation} \label{C Total}
C(x,t)=\frac{1}{\tau ^{AI} } \int _{0}^{t}C^{AI} (x,t) \, dt+C_{0} (x) ,
\end{equation}

\noindent where $C_{0} (x)$ is the initial distribution of
substitutionally dissolved impurity atoms at the initial stage of
annealing that can be approximated by the Gaussian distribution:

\begin{equation} \label{Gauss}
C_{0} (x)=C_{m} \exp \left[-\frac{\left(x-R_{p} \right)^{\, 2} }{2\Delta R_{p} ^{2} } \right] .
\end{equation}

Here $C_{m} $ is the maximal concentration of substitutionally
dissolved impurity.

Unfortunately, within the framework of the software for the
above-mentioned symbolic computation it is not possible to obtain
an explicit expression for integral \eqref{C Total} with $C^{AI}
(x,t)$ described by expression \eqref{Hom} or \eqref{Solution
Nonflow}. Therefore, the calculation of integral \eqref{C Total}
was carried out by an approximate numerical method using the
quadrature formula (Gauss 16-point formula)
\cite{Krylov-67,Davis-84}.

\section{Simulation of the interstitial redistribution of boron
implanted into silicon during low-temperature thermal treatments}
\begin{figure}[!ht]
\centering {
\begin{minipage}[!ht]{15.4 cm}
{\includegraphics[scale=1.2]{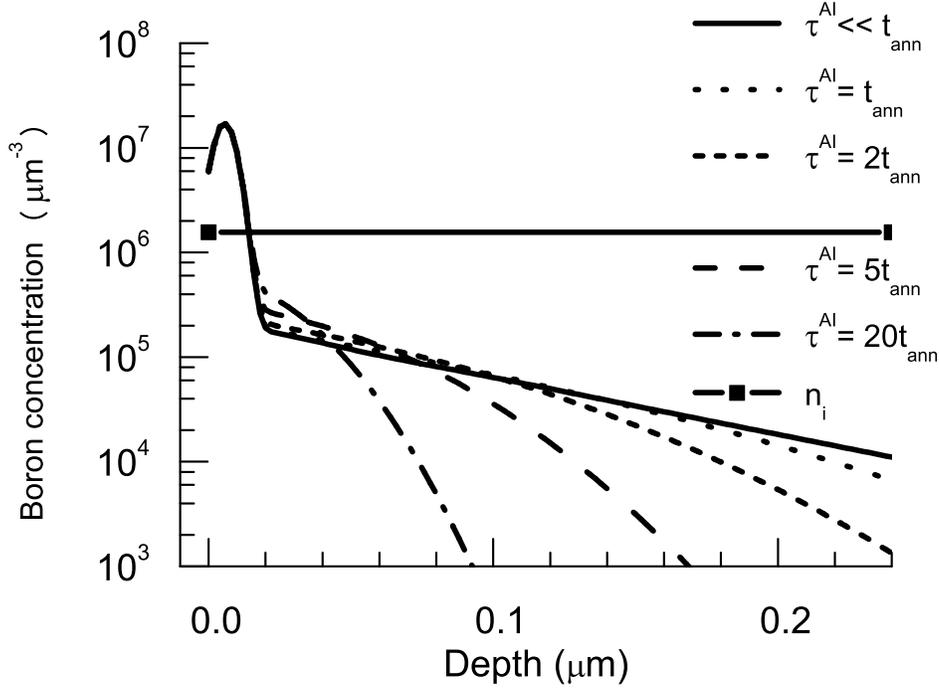}}
\end{minipage}}
\caption{Calculated boron concentration profiles after
low-temperature thermal treatment of implanted silicon substrate
for different ratios between the average lifetime $\tau ^{AI} $
and the duration of annealing $t_{ann}$} \label{fig:Profiles}
\end{figure}

In Fig.~\ref{fig:Profiles} the calculated profiles of
ion-implanted boron after low-temperature annealing are presented
which show the change in the shape of the low concentration
``tail'' formed due to the interstitial diffusion for different
ratios between the average lifetime $\tau ^{AI}$ and the duration
of thermal treatment $t_{ann} $. It is supposed that silicon is
implanted with 1 keV boron ions to a dose of
1.8$\times$10${}^{13}$ ion/cm${}^{2}$ ($R_{p} $ = 0.0056 $\mu $m;
$\Delta R_{p} $ = 0.0038 $\mu $m \cite{Burenkov-85}) and then
subjected to annealing with duration of 10 min at a temperature of
750 $^{\circ}$C. The average migration length of nonequilibrium
boron interstitials has been chosen to be equal to 0.08 $\mu $m.
It is clearly seen from Fig.~\ref{fig:Profiles} that if the
average lifetime of impurity interstitials $\tau ^{AI} $ is
significantly shorter than the annealing duration $t_{ann} $, the
``tail'' in the boron concentration profile represents a straight
line in the case of the logarithmic axis of concentration. With
increase in the average lifetime $\tau ^{AI}$, the shape of the
``tail'' becomes convex upwards, being characterized by an
increasing slope angle in the bulk of a semiconductor similar to
the Gaussian distribution. Thus, the calculations presented in
Fig.~\ref{fig:Profiles} show clearly that for nonzero initial
distribution of interstitial impurity atoms the ``tail'' in the
low concentration region of impurity distribution has the form of
the Gaussian distribution even for one event of impurity
interstitial migration if $\tau ^{AI} \ge t_{ann} $, whereas in
the case of continuous generation of impurity interstitials it
occurs only in the case of numerous events of interstitial
migration \cite{Cowern-90,Cowern-91}.

\begin{figure}[!ht]
\centering {
\begin{minipage}[!ht]{15.4 cm}
{\includegraphics[scale=1.2]{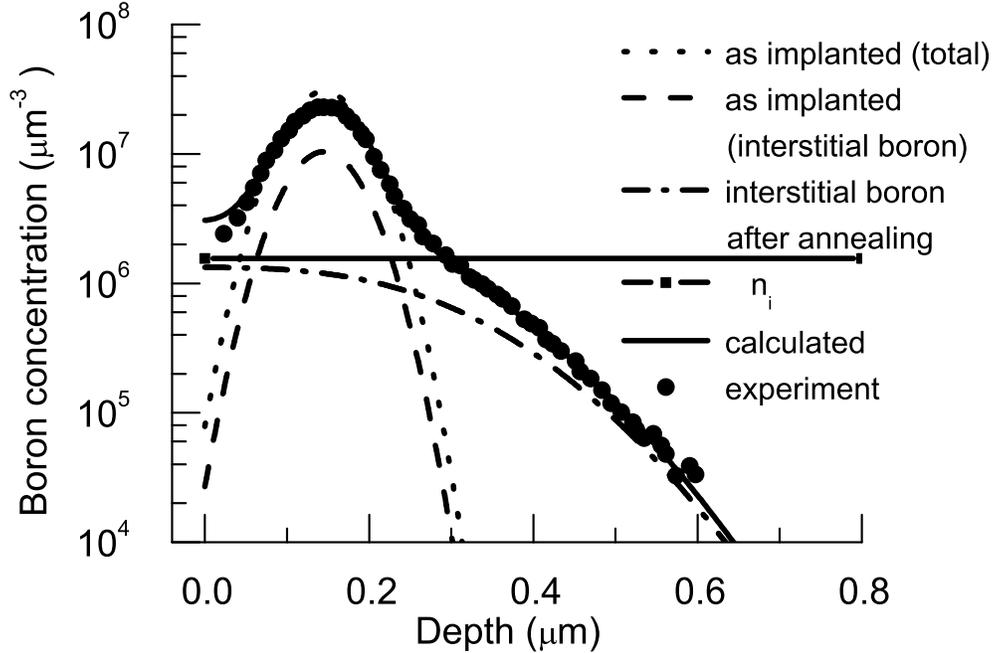}}
\end{minipage}}
\caption{Calculated boron concentration profile (solid line) after
thermal treatment of implanted silicon substrate at 750
$^{\circ}$C for 60 min. Gaussian distribution (dotted line - total
boron concentration and dashed line - concentration of
interstitial boron atoms) was used to approximate the initial
impurity profile. The dash-dotted curve represents interstitial
boron after annealing. The experimental data (impurity
distribution after annealing represented by filled circles) are
taken from Ref. \cite{Huang-97}} \label{fig:Huang}
\end{figure}

It follows from the experimental data that the ``tail'' in the
boron concentration profile after low-temperature treatment of
ion-implanted layer can have a shape of a straight line
\cite{Napolitani-99,Cristiano-04,Hamilton-07} as well as a convex
form characteristic for the Gaussian distribution
\cite{Huang-97,Hofker-75,Yeong-08,Jain-02} if the concentration
axis is logarithmic. For example, in Fig.~\ref{fig:Huang} the
boron concentration profile is presented which was measured in
\cite{Huang-97} by a method of the secondary ion mass spectroscopy
(SIMS). Both the p-type and n-type Czochralski grown (CZ)
commercial silicon wafers with conductivities of 4-6 $\Omega$cm
were used. Boron was implanted at room temperature with an energy
of 40 keV to a dose $Q$ of 3$\times$10${}^{14}$ cm${}^{-2}$. SIMS
profile of the implanted boron after annealing at 750 ${^\circ}$C
with duration $t_{ann}$ = 1 h is presented in
Fig.~\ref{fig:Huang}. by filled circles. It can be seen from
Fig.~\ref{fig:Huang} that the maximal boron concentration after
implantation $C_{m}$ = 3.04$\times$10$^{7}$ $\mu $m$^{-3}$ is near
the limit of boron solubility in silicon $C_{sol}$ =
2.33$\times$10$^{7}$ $\mu $m$^{-3}$ for a temperature of 750
$^{\circ}$C \cite{Solmi-01}. According to \cite{Haddara-00}, the
boron diffusivity of substitutionally dissolved atoms $D_{i}$ for
this temperature is equal to 2.658$\times$10$^{-10}$ $\mu
$m$^{2}$/s. Then, for the thermal treatment for 1 h the
characteristic diffusion length of impurity redistribution is
$L_{i} =\sqrt{D_{i} t_{ann} } $ = 9.78$\times$10${}^{-4}$ $\mu $m
or it is approximately equal to 1 nm. On the other hand, it
follows from Fig.~\ref{fig:Huang} that the actual $L_{i} $ is
approximately equal to 0.1 $\mu $m, i.e., 100 times longer. If we
suppose that boron diffusion occurs due to the formation,
migration, and dissociation of the ``impurity atom -- intrinsic
point defect'' pairs, it is necessary to accept that the average
concentration of nonequilibrium point defects responsible for
impurity diffusion is approximately 10$^{4}$ times higher than the
thermally equilibrium value. Not rejecting the possibility of such
a strong radiation-enhanced diffusion, we shall carry out modeling
of the redistribution of ion-implanted boron based on a more
natural mechanism of migration of nonequilibrium impurity
interstitials.

The following values of the model parameters were used to provide
the best fit of the calculated boron concentration profile to the
experimental one: \textbf{The parameters prescribing the initial
distribution of implanted boron:} $Q$ = 3.2$\times$10$^{14}$
cm$^{-2}$; $R_{p} $ = 0.145 $\mu $m; $\Delta R_{p} $ = 0.042 $\mu
$m; the fraction of the boron atoms in the interstitial position
$p^{AI} $ = 34.2 \%. \textbf{The parameters specifying the process
of interstitial diffusion:} the average migration length of boron
interstitials $l_{AI} $ = 0.11 $\mu $m. The reflecting boundary
condition is imposed on the surface of a semiconductor when
describing diffusion of boron interstitials. It is also supposed
that the average lifetime of nonequilibrium impurity interstitials
$\tau ^{AI} $ is equal to the duration of annealing $t_{ann} $ = 1
h. It is worth noting that the parameters prescribing the initial
distribution of implanted boron is approximately equal to the
values tabulated in \cite{Burenkov-85}\textbf{:} $R_{p} $ = 0.131
$\mu $m; $\Delta R_{p} $ = 0.0451 $\mu $m; $S_{k} $ = - 0.8 $\mu
$m; $R_{m} $ = 0.143 $\mu $m, where $S_{k} $ and $R_{m} $ are
respectively the skewness and the position of a maximum of
impurity distribution as implanted which is described by the
Pearson type IV distribution \cite{Burenkov-85}.

It can be seen from Fig.~\ref{fig:Huang} that there is an
excellent agreement between the calculated total boron
concentration profile and boron profile measured by SIMS. A little
difference in the near surface region follows from the assumption
about the reflecting boundary for boron interstitials. In point of
fact part of impurity interstitials cross over the surface of a
semiconductor.

\begin{figure}[!ht]
\centering {
\begin{minipage}[!ht]{15.4 cm}
{\includegraphics[scale=1.2]{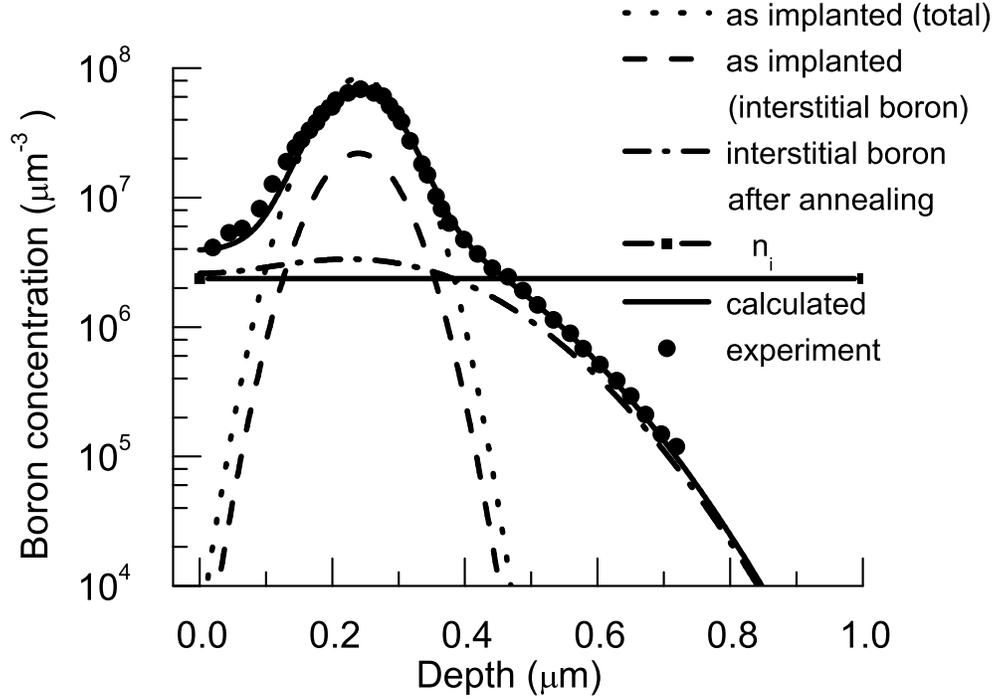}}
\end{minipage}}
\caption{Calculated boron concentration profile (solid line) after
thermal treatment of implanted silicon substrate at 800
$^{\circ}$C for 35 min. Gaussian distribution (dotted line - total
boron concentration and dashed line - concentration of
interstitial boron atoms) was used to approximate the initial
impurity profile. The dash-dotted curve represents interstitial
boron after annealing. The experimental data (impurity
distribution after annealing represented by filled circles) are
taken from Ref. \cite{Hofker-75}} \label{fig:Hofker}
\end{figure}

The modeling results of interstitial diffusion of ion-implanted
boron during a low-temperature annealing at a temperature of 800
$^{\circ}$C for 35 min are presented in Fig.~\ref{fig:Hofker}. The
experimental data of \cite{Hofker-75} were used for comparison. In
the Ref. \cite{Hofker-75} boron was implanted with an energy of 70
keV to a dose $Q$ of 10${}^{15}$ cm${}^{-2}$. For the energy used
in ion implantation, the parameters of the Pearson type IV
distribution are: $R_{p} $ = 0.219 $\mu $m; $\Delta R_{p} $ =
0.0606 $\mu $m; $S_{k} $ = -0.9 \cite{Burenkov-85}. It is worth
noting that for asymmetry $S_{k} $ = -0.9 the position of a
maximum of boron concentration $R_{m} $ = 0.236 $\mu $m is close
to the value $R_{p} $ = 0.219 $\mu $m. It means that, just as in
the previous case, one can neglect the asymmetry of impurity
distribution and describe the boron profile after implantation,
including the distribution of boron atoms in the interstitial
position, by the symmetric Gaussian distribution \eqref{Initial
condition} \cite{Burenkov-80}.

The diffusivity of the substitutionally dissolved boron atoms
$D_{i} $ is equal to 1.928$\times$10${}^{-9}$ $\mu $m$^{2}$/s for
a temperature of 800 $^{\circ}$C \cite{Haddara-00}. Then, the
characteristic diffusion length of impurity redistribution $L_{i}
=\sqrt{D_{i} t_{ann} } $ = 2.01$\times$10${}^{-3}$ $\mu $m or
approximately 2 nm for thermal treatment during 35 min. On the
other hand, it can be seen from Fig.~\ref{fig:Hofker} that the
actual value of $L_{i}$ is approximately 0.1 $\mu $m, i.e., 50
times longer. To provide this value of $L_{i} $, the time-average
concentration of the nonequilibrium point defects responsible for
the impurity diffusion should be 2.5$\times$10$^{3}$ times higher
than a thermally equilibrium value. It is necessary to note that
in the case of radiation-enhanced diffusion of impurity atoms, due
to the formation, migration, and dissociation of ``impurity atom
-- intrinsic point defect'' pairs, all impurity atoms in the
``tail'' region should be substitutionally dissolved, i.e., to be
electrically active. However, the experimental data of
\cite{Hofker-75} show that the significant fraction of boron atoms
is electrically inactive in the ``tail''. This phenomenon is
easily explained within the framework of the interstitial
diffusion mechanism described above if the lifetime of the
interstitial boron atoms $\tau ^{AI}$ is near or longer the
annealing duration $t_{ann}$. Indeed, it follows from the later
condition that a significant part of boron atoms are staying in
the interstitial position, i.e. they are electrically neutral.

The following values of the model parameters were used to provide
the best fit of the calculated boron concentration profile to the
experimental one (see Fig.~\ref{fig:Hofker}): \textbf{The
parameters prescribing the initial distribution of implanted
boron:} $Q$ = 1.12$\times$10${}^{15}$ cm${}^{-2}$; $R_{p} $ = 0.24
$\mu $m; $\Delta R_{p} $ = 0.054 $\mu $m; the fraction of the
boron atoms in the interstitial position $p^{AI} $ = 26.5 \%. It
can be seen in Fig.~\ref{fig:Hofker} that the maximal boron
concentration after implantation $C_{m} $ = 8.3$\times$10${}^{7}$
$\mu $m$^{-3}$ is more than 2 times higher than the limit of boron
solubility in silicon $C_{sol}$ = 3.431$\times$10${}^{7}$ $\mu
$m$^{-3}$ for a temperature of 800 $^{\circ}$C \cite{Solmi-01}.
\textbf{The parameters specifying the process of interstitial
diffusion:} the average migration length of boron interstitials
$l_{AI} $ = 0.14 $\mu $m. As in the previous simulation, the
reflecting boundary condition is imposed on the surface of a
semiconductor when describing diffusion of boron interstitials.
The average lifetime of nonequilibrium impurity interstitials
$\tau ^{AI} $ = 48,3 min is chosen to be longer than the duration
of annealing $t_{ann}$ = 35 min. It is worth noting that the
parameters prescribing the initial distribution of implanted boron
is approximately equal to the values listed in the
above-mentioned tables \cite{Burenkov-85}.

It can be seen from Fig.~\ref{fig:Hofker} that there is excellent
agreement between the calculated total boron concentration profile
and boron profile measured by SIMS. It is interesting to note that
the concentration of interstitial boron atoms is approximately
equal to the total boron concentration at the end of the ``tail''.
It means that this part of the ``tail'' is electrically inactive
that is also consistent with the experimental data
\cite{Hofker-75}.

This paper considers the diffusion of boron interstitial atoms
generated either during ion implantation or at the initial stage
of annealing. In this meaning the case of interstitial diffusion
under consideration is opposite to the investigations of
\cite{Cowern-90,Cowern-91,Velichko-07,Velichko-2011} where
continuous generation of impurity interstitials is supposed. It
follows from the simulation results obtained (see Figs. 2 and 3)
that the model of diffusion of previously generated boron
interstitials allows one to explain the formation of ``tails'' in
nonamorphized silicon layers during low-temperature annealing with
duration of 1 h or shorter. On the other hand, some increase in
the boron concentration in the ``tail'' region occurs for
annealing duration of 21 h at 800 $^{\circ}$C \cite{Hofker-75} in
comparison with the simulated boron profile for 1 h annealing
presented in Fig.~\ref{fig:Hofker}. We suppose that the additional
generation of boron interstitials for such long-time treatments
occurs due to the annealing or rearrangement of clusters or
radiation defects.

\section{Conclusions}
To explain the redistribution of ion implanted boron during
low-temperature annealing of nonamorphized silicon layers, a model
of interstitial impurity migration is proposed. In contrast to
other models of interstitial diffusion, it is supposed that
nonequilibrium boron interstitials are generated either during ion
implantation or at the initial stage of annealing and migration
inward and to the surface of a semiconductor during the basic
stage of annealing. It was shown that such kind of interstitial
diffusion results in a ``tail'' in the boron concentration profile
which has a form of a straight line for the logarithmic
concentration axis if the average lifetime of impurity
interstitials $\tau ^{AI} $ is significantly shorter than
annealing duration $t_{ann} $ and becomes convex upwards similar
to the Gaussian distribution with increasing $\tau ^{AI} $.

The calculated impurity concentration profiles are in excellent
agreement with the experimental data that describe redistribution
of implanted boron for low-temperature annealing at 750
$^{\circ}$C for 1 h and at 800 $^{\circ}$C for 35 min if $\tau
^{AI} $ is equal to 1 h and to 48.3 min, respectively, i.e., the
average lifetime of boron interstitials is equal or rather longer
than the duration of annealing. The average migration length of
impurity interstitial atoms is equal to 0.11 $\mu$m and to 0.14
$\mu$m for the annealing at temperatures 750 and 800 $^{\circ}$C,
respectively. The simulation results also show that 34.2 \% and
26.5 \% of the implanted boron atoms occupy interstitial position
at the initial stage of annealing.

It is worth noting that the experimental phenomenon of incomplete
electrical activation of boron atoms in the ``tail'' region is
naturally explained within the framework of the model proposed.
This fact and the excellent agreement with the measured boron
profiles give evidence in favor of the model considering the
formation of the main part of boron interstitials generated either
during ion implantation or at the initial stage of annealing if we
deal with low-temperature treatment of nonamorphized silicon
layers.

\end{document}